\documentclass[twocolumn]{aastex631}
\usepackage{graphicx} 
\usepackage{epstopdf}
\usepackage{amsmath,soul} 
\usepackage{placeins}

\newcommand\bb[1]{\mbox{\boldmath{$#1$}}}
\newcommand\grad{\bb{\nabla}}
\newcommand\bcdot{\,\bb{\cdot}\,}
\newcommand\bdbldot{\,\bb{:}\,}
\newcommand\btimes{\,\bb{\times}\,}
\newcommand\rmd{\mathrm{d}}
\newcommand\rms{\mathrm{s}}

\begin{document}

\title{First-principles Measurement of Ion and Electron Energization in Collisionless Accretion Flows}

\author[0000-0001-9073-8591]{Evgeny A.\ Gorbunov}
\affiliation{Centre for mathematical Plasma Astrophysics, Department of Mathematics,
KU Leuven, Celestijnenlaan 200B, B-3001 Leuven, Belgium}

\author[0000-0002-7526-8154]{Fabio Bacchini}
\affiliation{Centre for mathematical Plasma Astrophysics, Department of Mathematics,
KU Leuven, Celestijnenlaan 200B, B-3001 Leuven, Belgium}
\affiliation{Royal Belgian Institute for Space Aeronomy, Solar-Terrestrial Centre of Excellence, Ringlaan 3, 1180 Uccle, Belgium}

\author[0000-0003-3816-7896]{Vladimir Zhdankin}
\affiliation{Department of Physics, University of Wisconsin-Madison, Madison, WI 53706, USA}

\author[0000-0001-9039-9032]{Gregory R.\ Werner}
\affiliation{Center for Integrated Plasma Studies, Department of Physics, University of Colorado, 390 UCB, Boulder, CO 80309-0390, USA}
\author[0000-0003-0936-8488]{Mitchell C.\ Begelman}
\affiliation{JILA, University of Colorado and National Institute of Standards and Technology, 440 UCB, Boulder, CO 80309-0440, USA}
\affiliation{Department of Astrophysical and Planetary Sciences, University of Colorado, 391 UCB, Boulder, CO 80309-0391, USA}
\author[0000-0001-8792-6698]{Dmitri A.\ Uzdensky}
\affiliation{Center for Integrated Plasma Studies, Department of Physics, University of Colorado, 390 UCB, Boulder, CO 80309-0390, USA}
\affiliation{Rudolf Peierls Centre for Theoretical Physics, Clarendon Laboratory, University of Oxford, Parks Rd., Oxford, OX1 3PU, UK}

\correspondingauthor{Evgeny A. Gorbunov}
\email{evgeny.gorbunov@kuleuven.be}

\begin{abstract}
We present the largest 3D Particle-in-Cell shearing-box simulations of turbulence driven by the magnetorotational instability, for the first time employing the realistic proton-to-electron mass ratio. We investigate the energy partition between relativistically hot electrons and subrelativistic ions in turbulent accreting plasma, a regime relevant to collisionless, radiatively inefficient accretion flows around supermassive black holes such as those targeted by the Event Horizon Telescope. We provide a simple empirical formula to describe the measured heating ratio between ions and electrons, which can be used for more accurate global modeling of accretion flows with standard fluid approaches such as general-relativistic magnetohydrodynamics.
\end{abstract}

\section{Introduction}
It is generally believed that in radiatively inefficient accretion flows (RIAFs, e.g.,\citealt{ichimaru1977,rees1982,Narayan_1994_apjl,Naryan_1995_apj_a,Narayan_1995_apj_b,Narayan_1995_nat,quataert_2003,Yuan_2014}) around astrophysical compact objects, the typical accretion time is much faster than the average collision time between plasma particles (i.e.,\ electrons and protons; e.g.,\ \citealt{Mahadevan_1997,Yuan_2014}); in these environments, collisionless plasma physics may play a dominant role in defining the energetics of the accretion flows. In such a regime, Coulomb collisions cannot efficiently equilibrate the temperature of ions and electrons, thus leading to the existence of a two-temperature state \citep[e.g.,][]{shapiro_1976,rees1982,Blackman_1999,Kawazura_2019,Zhdankin_2019} 
and non-thermal  distributions of particle species \citep[e.g.,][]{zhdankin_etal_2017,Comisso_2018,Zhdankin_2018,comisso_2019,Zhdankin_2019,Zhdankin_2021}. 
Electrons in these environments might also become relativistically hot, and radiate rapidly \citep[e.g.,][]{Yuan_2014}. This situation applies to the surroundings of many supermassive black holes (SMBHs) in active galactic nuclei (AGNs), including the current targets of observational campaigns such as the Event Horizon Telescope (EHT), i.e.,\ M87$^*$ and SgrA$^*$ (\citealt{EHT2019e,EHT2022e}). In such systems, the magnetorotational instability (MRI;  \citealt{balbushawley1991,balbushawley1998}) could represent a simple and powerful mechanism promoting outward angular-momentum transport (AMT) and acting as a turbulence driver. Even for magnetically dominated disks, where one would expect axisymmetric MRI modes to be suppressed \citep[e.g.,][]{Porth_2021}, nonaxisymmetric MRI modes (generally termed superalfv\'enic rotational instabilities or SARIs) can efficiently develop to drive turbulence throughout the disk \citep{begelman2022,Goedbloed_2022,brughmans2024}. This is especially relevant considering that recent 3D global general-relativistic magnetohydrodynamic (GRMHD) simulations of accretion \citep{Ripperda_2022} have shown that realistic disks are likely inhomogeneous in terms of magnetization so that large-scale, strongly magnetized portions of the disk coexist with regions where the magnetization is much lower and the MRI could develop unimpeded (see also \citealt{Spruit_2005}).

GRMHD simulations are the de-facto standard approach to model the global structure of SMBH accretion disks (e.g.,\ \citealt{McKinney_2004,Fragile_2007,Moscibrodzka_2009,Tchekhovskoy_2010,Naryan_2019,Ressler_2015,Moscibrodzka_2016,Moscibrodzka_2017,Chael_2018,Ripperda_2019,Dexter_2020,Ripperda_2020,Narayan_2022,Ripperda_2022,Scepi_2022,Scepi_2023}).
The most widely applied GRMHD models treat plasmas as a quasi-neutral, collision-dominated single fluid, without distinguishing electron and ion dynamics. To obtain emission spectra and radiation maps from GRMHD simulations, knowledge of the electron temperature $T_e$ (or ion-to-electron temperature ratio $T_i/T_e$) is required, as the electrons are believed to produce most of the observed radiation; however, this quantity is not intrinsically available in GRMHD models. Early modeling efforts simply set the temperature ratio to a constant \citep{Moscibrodzka_2009,Moscibrodzka_2016,Moscibrodzka_2017} in the bulk of accretion disks. Later works evolved the equations governing ions and electron thermodynamics separately, allowing the self-consistent mapping of the electron energy density (i.e.,\ electron temperature, \citealt{Ressler_2015,Chael_2018,Dexter_2020}). These evolution equations rely on sub-grid-scale heating prescriptions to model the dissipation of energy at kinetic ion and electron scales, which are by definition not captured by GRMHD.

The past few years have seen significant effort devoted to obtaining heating prescriptions (or, in other words, the ion-to-electron energy partition) in collisionless plasmas using results from kinetic theory, which may better describe SMBH accretion flows.
The aforementioned subgrid heating prescriptions {were obtained from nonrelativistic gyrokinetic (GK) theory \citep[e.g.,][]{Howes_2010}, and numerically from hybrid GK, with fluid electrons and kinetic ions \citep[e.g.,][]{Kawazura_2019}. Later works using hybrid GK \citep{,kawazura_etal_2020} and nonrelativistic reduced MHD \citep{Kawazura_2022}} suggested that at small scales, where compressive and Alfv\'enic turbulent cascades decouple, ions are preferentially heated over electrons via compressive fluctuations. They also argued that in MRI turbulence the power of compressive fluctuations is significantly larger than that of Alfv\'enic fluctuations \citep{Kawazura_2022} for the case of a strong azimuthal (i.e.,\ toroidal) mean magnetic field threading an accretion disk. 
In addition, fully kinetic Particle-in-Cell (PIC) simulations of isolated relativistic magnetic reconnection (without turbulence) parametrized the heating ratio with respect to the plasma magnetization \citep{Werner_2017}. Finally, a series of externally driven relativistic-turbulence PIC simulations \citep[e.g.,][]{Zhdankin_2019,Zhdankin_2021} revealed that the type of driving (solenoidal or compressive) influences the resulting energy partition between ions and electrons, with compressive driving providing more energy to ions.
In this context, a fully consistent investigation of realistically driven turbulent, collisionless plasma accretion (down to the kinetic scales) characterizing the ion-to-electron heating ratio is still missing.
The kinetic modeling of MRI-driven turbulence has the potential to address the problem of energy partition in accretion disks around black holes without ad-hoc prescriptions for the type of turbulent driving present in such systems, while also capturing electron and ion dynamics up to the relativistic energies expected in these environments.  

In this work, we perform the first 3D fully kinetic investigation of MRI-driven turbulence in accreting ion--electron plasmas with a realistic mass ratio, and measure the subsequent heating ratio.
We model a small sector of the accretion disk using the shearing-box (SB) approach \citep[][]{stone_1996,balbushawley1998,Sano_2004}, a method that was previously applied in fluid modeling of accretion flows \citep[e.g.,][]{Sharma_2006,Stone_2010,Pastorello_2013,Hirai_2018,Kawazura_2024} as well as in pioneering 2D kinetic simulations, hybrid approaches (with fluid electrons), pair plasmas, and/or small mass ratios \citep[e.g.,][]{Riquelme_2012,hoshino2013,hoshino2015,Kunz_2016,inchingolo2018,Bacchini_2022,bacchini_2024,Sandoval_2024}. As we will demonstrate, our novel three-dimensional ion--electron SB simulations are capable of producing an MHD-like turbulent cascade and retrieving AMT characteristics consistent with fluid models, while allowing us to assess the ion-to-electron heating ratio from first principles. 

\section{Simulation setup}
\subsection{Kinetic shearing box with orbital advection}
The simulations were conducted with the PIC code \textsc{Zeltron} \citep{Cerutti_2013}, using our kinetic shearing box with orbital advection (KSB-OA; \citealt{Bacchini_2022,bacchini_2024}) paradigm. We model a small Cartesian box defined with $(x,y,z)$ coordinates, with the $x$-coordinate being the local radial direction (i.e., with the total distance from the central object being $r = x + R_0$, where $R_0$ is the distance between the center of the simulation box and the center of rotation and $|x| \ll R_0$), $y$ being the toroidal coordinate, and $z$ being the vertical coordinate. The simulation domain rotates around the central object with orbital frequency~$\Omega_0$. A background, linearized Keplerian shear with velocity profile $\bb{v}_\rms (x) = -s\Omega_0 x \widehat{\bb{e}}_y$ is present such that $v_\rms \ll c$, where $s= 3/2$ and $c$ is the speed of light.
In the KSB-OA formalism, the electric field and particle momenta are computed in a frame comoving with the shear velocity $\bb{v}_\rms$. In contrast, the magnetic field is computed in the lab frame \citep{Bacchini_2022,bacchini_2024}. The resulting evolution equations for the magnetic field $\bb{B}$ and the (comoving) electric field $\bb{E}$ are 
\begin{align}
    \partial_t \bb{B} =& - c \grad\btimes\left(\bb{E} - \frac{\bb{v}_{\rm s}}{c}\btimes\bb{B}\right),\\
    \partial_t \bb{E} =& c\grad\btimes\bb{B} - 4\pi\bb{J}  \nonumber\\
    &- \bb{v}_\rms\btimes\left(\grad\btimes\bb{E}\right)+\frac{\bb{v}_\rms}{c}\btimes\left(\grad\btimes\left(\frac{\bb{v}_\rms}{c}\btimes\bb{B}\right)\right),
\end{align}
where $\bb{J}$ is the comoving electric current computed in the lab frame. 
The evolution equation for the spatial part of the comoving 4-velocity $\bb{u}$ for a particle of mass $m$, charge $q$, and Lorentz factor $\gamma = \sqrt{1+u^2/c^2}$ in the lab frame is
\begin{align}
    & \frac{\rmd\bb{u}}{\rmd t} = \frac{q}{m}\left(\bb{E} + \frac{\bb{u}}{c\gamma}\btimes\bb{B}\right) + 2\bb{u}\btimes\mathbf{\Omega}_0 + s\Omega_0 u_{x}\widehat{\bb{e}}_y, \label{eq:momenta}
\end{align}
which includes the action of the Lorentz force, the Coriolis force (with $\bb{\Omega}_0=\Omega_0\widehat{\bb{e}}_z$), and the radial component of the gravitational force. The particle coordinate equation in the lab frame is obtained via a Galilean transformation of the comoving momentum to the lab frame, 
\begin{align}
    & \frac{\rmd\bb{x}}{\rmd t} = \frac{\bb{u}}{\gamma} + \bb{v}_{\rm s}(\bb{x}). 
\end{align}
The boundaries in the $y$- and $z$-directions are periodic, while along $x$ shearing-periodic boundary conditions are applied\footnote{For more details on the numerical implementation, see \citealt{Bacchini_2022}.}.

\subsection{Simulation parameters}
The height of the simulation box is chosen to be $L_z = 2\lambda_{\rm MRI}$, with the most-unstable MRI wavelength $\lambda_{\rm MRI}$ depending on the initial nonrelativistic Alfv\'en speed $v_{\rm A,0}$ (defined with respect to ions, see below) and $\Omega_0$ as $\lambda_{\rm MRI} \approx 2\pi v_{\rm A,0}/\Omega_0$ \citep{balbushawley1998}. 
Like in previous works \citep{Bacchini_2022,bacchini_2024}, the box aspect ratio was chosen to be $L_y = 2 L_x = 4 L_z$, to allow the development of parasitic instabilities on top of the primary MRI \citep{goodmanxu1994,pessahgoodman2009}. Parasitic modes have been shown to be essential for the transition to a turbulent regime during the late stages of MRI evolution \citep{Bacchini_2022} and therefore must be included.

In this work, we model collisionless ion--electron plasma $q_i = -q_e = q$ with a realistic proton-to-electron mass ratio $m_i/m_e = 1836$. We focus on the \emph{semirelativistic} regime \citep{Werner_2017},
which is characterized by electrons being relativistically hot ($\theta_e \equiv T_e/(m_e c^2)\ge1$), while ions are initially cold ($\theta_i \equiv T_i/(m_i c^2) = 1/96 \le 1$). By setting the two plasma species to have equal temperatures $T_i=T_e$ at the start of the simulation, we obtain an initial dimensionless electron temperature $\theta_e = (m_i/m_e) \theta_i \approx 19$. With this choice of parameters, the average Lorentz factor of electrons is $\gamma_e \approx 3\theta_e \approx 57$, and the Lorentz factor of ions is $\gamma_i \approx (3/2)\theta_i + 1 \approx 1.016$ at initialization. Due to particle energization, the ions may enter the relativistic regime during the late stages of MRI turbulence. Particle velocities are initially sampled from a Maxwell-J\"uttner distribution, and particles are uniformly distributed in space with number density $n_i = n_e = n$. Initially, a weak magnetic field $\bb{B}_0 = (\delta B_x,0, B_z)$ is present in the system. To save computational time (and because we are primarily interested in the nonlinear MRI stage), we include a small-amplitude initial perturbation $\delta B_x = 0.2 B_z \sin(2\pi z/ L_z)$ to trigger an earlier onset of the instability. The $B_z$ component of the initial magnetic field is defined via the ratio of Alfv\'en speed to the speed of light $v_{{\rm A},0}/c = B_z / \sqrt{4\pi m_i n c^2} = 0.008$, which we set $\ll 1$ to respect the intrinsic SB assumptions of nonrelativistic conditions \citep{Bacchini_2022}. The numerical grid spacing in our simulations is equal to the relativistic electron inertial length\footnote{As plasma heats up during the simulation, the electron inertial length increases, leading to electron scales being better resolved during the late MRI stages.}, $\Delta x = \Delta y = \Delta z = d_{e,0}^{\rm R} = c/\sqrt{4\pi n q^2/(\gamma_e m_e)}$.

The characteristic size of our simulation can be defined via the scale separation between $\lambda_{\rm MRI}$ and the largest kinetic scale, i.e.,\ the average ion Larmor radius $\rho_i$, initially defined via its nonrelativistic expression $\rho_{i,0} \equiv \sqrt{\theta_i} m_i c^2/(q B_0)$. The ratio $\lambda_{\rm MRI}/\rho_{i,0}$ can be expressed in terms of the ratio between the initial ion cyclotron frequency $\omega_{{\rm c},i} = qB_0/(m_ic)$ and the orbital frequency~$\Omega_0$. For our simulation, we set\footnote{{In real SMBH RIAFs, one expects $\omega_{\rm c,i}/\Omega_0 \sim 10^7\mbox{--}10^9$.}} $\omega_{\rm c,i}/\Omega_0 = 8$ which implies that $\lambda_{\rm MRI}/\rho_{i,0} \approx 4$ at initialization. Since electrons are relativistically hot, their initial Larmor radius $\rho_{e,0} \equiv m_e c^2\sqrt{\gamma_e^2 - 1} / (qB_0)$ is significantly larger than its nonrelativistic value, setting the ion--electron gyroradius ratio to be $\rho_{i,0}/\rho_{e,0} \approx 3.3$. With these parameters, the initial ion Larmor radius to relativistic electron inertia length ratio is $\rho_{i,0}/d_{e,0}^{\rm R}\approx72$, the initial magnetization is $\sigma_0 = 6\times10^{-5}$, and the initial plasma beta is $\beta_0 = 620$.
These choices require the use of a large total number of cells $N_{\rm cells} = N_x \times N_y \times N_z = 1152 \times 2304 \times 576$. The number of particles per cell per species $N_{\rm ppc} = 27$. This is the largest simulation performed for this work, and we progressively reached it with a series of simulations with $\omega_{{\rm c}, i}/\Omega_0 = 2,4,6,8$. The simulations were performed for $t=15P_0 = 30\pi / \Omega_0$, where $P_0$ is the orbital period.

\section{Results}
\subsection{Turbulence evolution and properties}
\begin{figure*}[t]
    \centering 
    \includegraphics[width=1\textwidth]{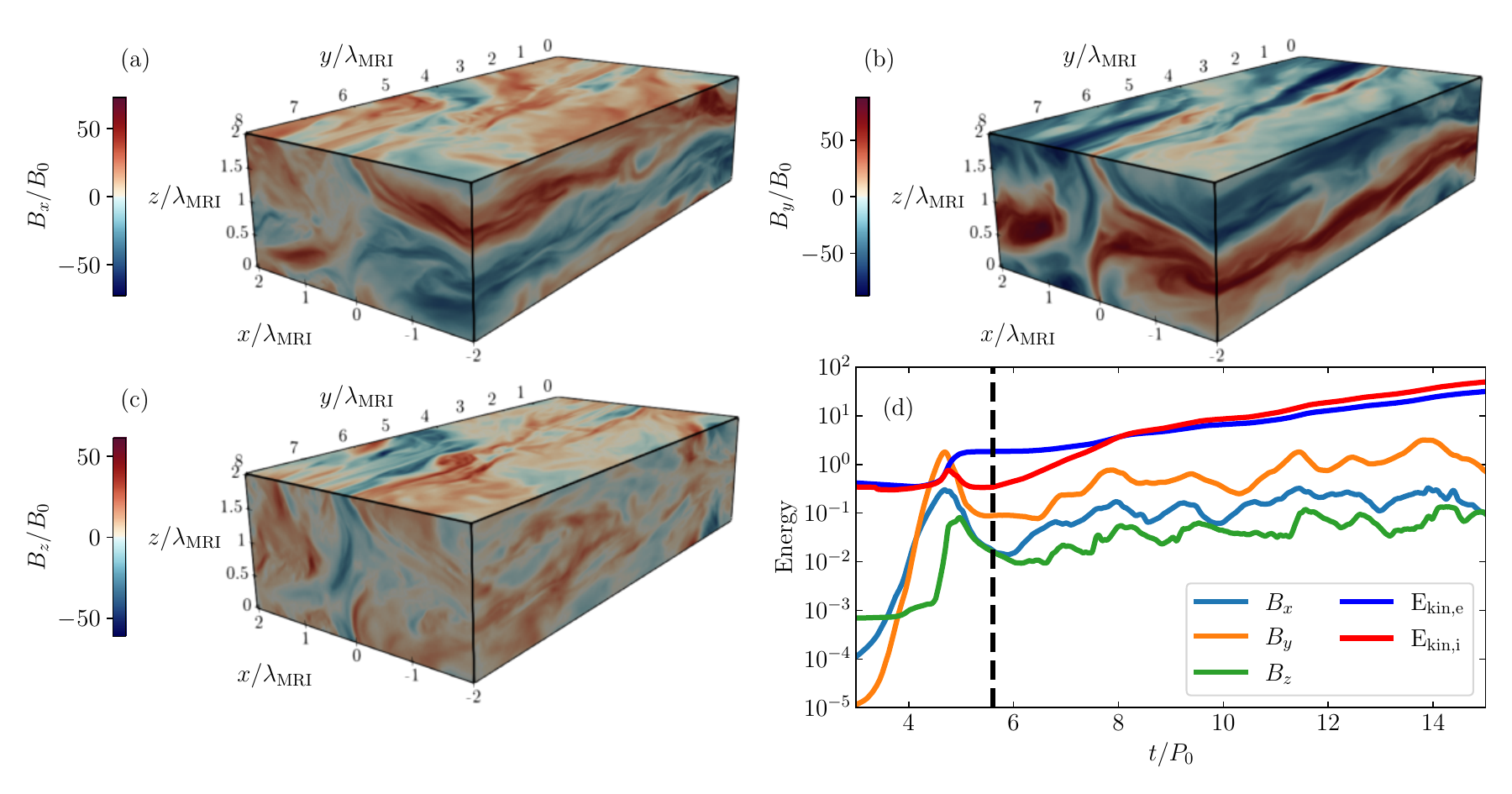}
    \caption{Structure of the (a) $B_x$, (b) $B_y$ and (c) $B_z$ magnetic-field components during the turbulent stage, at $t = 12P_0$; (d) Energy evolution of magnetic-field components and kinetic energies of ions and electrons, normalized by the total energy at $t=0$. The dashed line depicts a moment after which the RR force was turned off.}
    \label{fig:fields}
\end{figure*}
\begin{figure*}
    \centering 
    \includegraphics[width=1\textwidth]{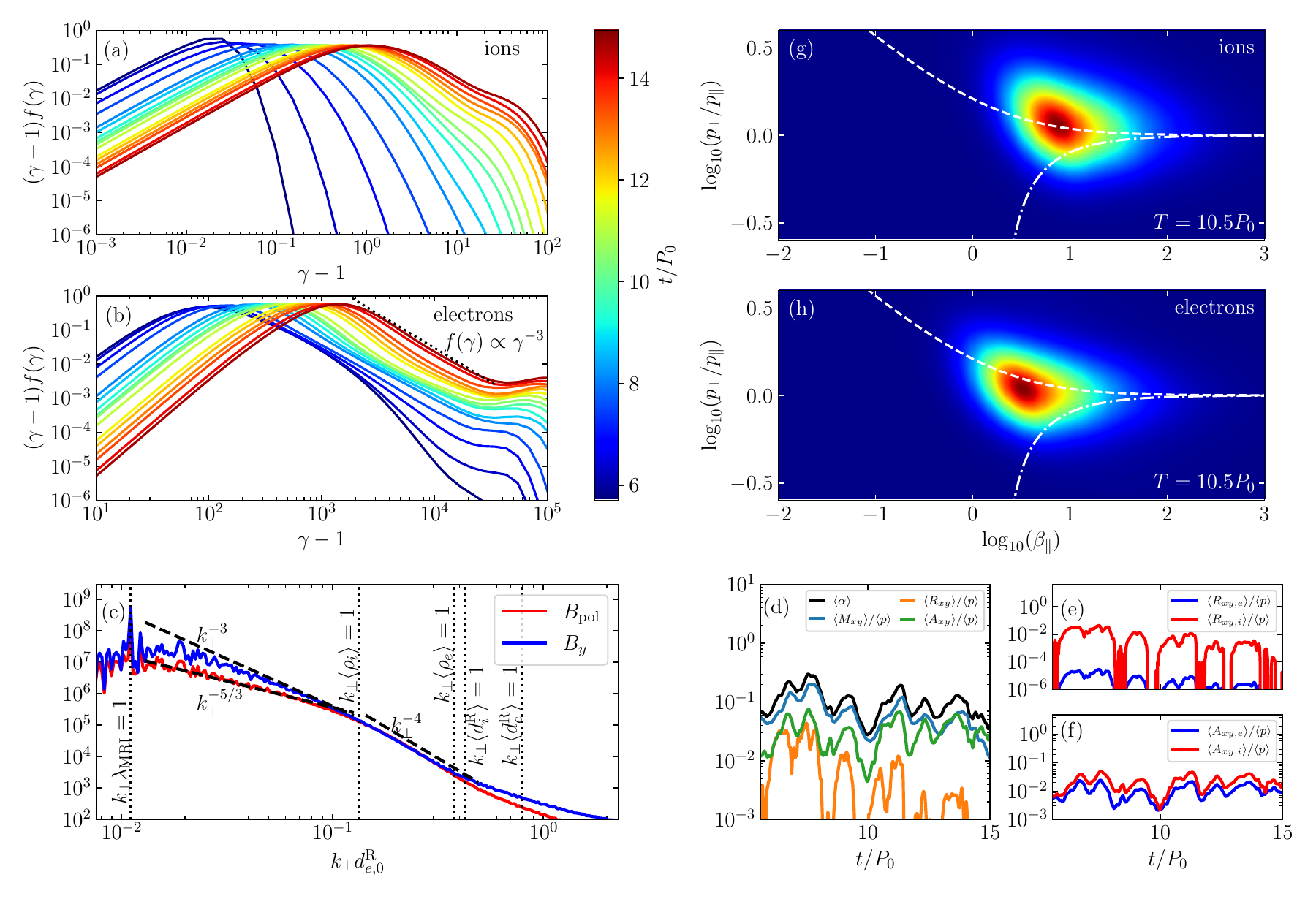}
    \caption{Different MRI-driven electron--ion turbulence properties. (a,b): Particle energy distribution functions during the turbulence stage; (c): Turbulent spectra for the poloidal magnetic field $B_{\rm pol} \equiv \sqrt{B_x^2+B_z^2}$ and the toroidal magnetic field $B_y$, averaged over $t = 10.5P_0\mbox{--}12.5P_0$. Perpendicular wavenumbers are normalized by the initial relativistic electron skin depth $d^{\rm R}_{e,0}$ (d): Temporal evolution of the total $\langle\alpha\rangle$-parameter (black line) and total stresses, normalized by the average total pressure~$\langle p \rangle$. (e,f): Reynolds and anisotropic stresses for different species; (g,h): Distribution of $(p_{\perp}/p_{\parallel},\beta_{\parallel})$ for (g) electrons and (h) ions at $t=10.5P_0$. The approximate firehose ($p_{\perp,j}/p_{\parallel,j} > 1 - 2/\beta_{\parallel,j}$)  and mirror instability ($p_{\perp,j}/p_{\parallel,j} > 1/2 (1 + \sqrt{1 + 4/\beta_{\parallel,j}})$) thresholds are indicated in dash-dotted and dashed lines, respectively. }\label{fig:figure2}
\end{figure*}

The ion--electron MRI evolution in our runs follows the general trends of kinetic MRI observed in previous works \citep[e.g.,][]{Riquelme_2012,hoshino2015,Bacchini_2022,bacchini_2024}. From the initial conditions, the weak poloidal field $\bb{B}_{\rm pol}= \{B_x,B_z\}$ bends and stretches, producing a toroidal $B_y$ component. Throughout the linear stage, the MRI forms two horizontal channel flows of size $\sim\lambda_{\rm MRI}$ in~$z$, consistent with theoretical MHD expectations \citep[e.g.,][]{goodmanxu1994}. The magnetic field in these channels grows exponentially with a growth rate predicted by linear theory \citep{balbushawley1991}. The linear growth in our simulation, preceded by the pre-instability stage (i.e., before the initial magnetic field becomes unstable), occurs during $t\approx 3.5P_0\mbox{--}4.5P_0$.
The growth of the magnetic-field amplitude inside the channels leads to the thinning of the large-scale current sheets located at the channel interfaces. This process continues until the onset of tearing and drift--kink instabilities, which disrupt the channels and drive large-scale reconnection \citep{Bacchini_2022}, with this stage occurring at $t\approx4.5P_0\mbox{--}5.0P_0$. Finally, after $t\approx5.0P_0$, the system exhibits turbulent 3D dynamics. We focus on this stage, potentially representative of the typical state of realistic accretion flows, to measure plasma heating. In this turbulent regime, channels may occasionally reform, to be later disrupted again by reconnection; such cycles can occur repeatedly throughout the nonlinear MRI stage, and represent the natural behavior of MRI turbulence \citep{Bacchini_2022}. The spatial distribution of the magnetic field at a representative moment during the turbulent stage ($t=12P_0$) is shown in Fig.~\ref{fig:fields}(a--c). The energy evolution during the simulation is given in Fig.~\ref{fig:fields}(d).

The violent energy release from the reconnecting magnetic field during channel disruption, prior to the turbulent stage, poses a particular problem that has to be addressed before the turbulent plasma heating can be measured. This magnetic energy goes into both ions and electrons, preferentially heating the former \citep{Werner_2017}. If not dealt with, this sudden growth in kinetic energy will lead to two issues: (i) the ions can develop a nonthermal tail in their distribution before the onset of turbulence (which is not expected in the semirelativistic regime), and (ii) a significant portion of the ions become relativistic, therefore impacting the turbulence properties. As our previous work has shown, initially thermal electrons can obtain a nonthermal tail during the turbulent stage \citep{bacchini_2024} solely via the action of MRI turbulence, and our main concern here is therefore related to avoiding nonthermal particle acceleration (NTPA) of ions \textit{prior} to reaching the turbulent stage. While previous works on relativistic turbulence have shown that compressive driving in collisionless turbulence can produce nonthermal populations of nonrelativistic ions \citep{Zhdankin_2019}, while solenoidal driving may not, and both types of driving can lead to NTPA for relativistic ions \citep{Zhdankin_2021}, 
it remains to be seen whether or not MRI turbulence is capable of effective production of nonthermal ions in the semirelativstic regime studied in this paper. Therefore, we wish to suppress ion NTPA during the preturbulence stages of MRI, making the turbulent stage agnostic of the initial linear growth stage and follow-up reconnection. In this way, the nonlinear stage can start with subrelativistic, quasithermal ions.

To extinguish undesirable NTPA during the preturbulence stages, we employ the strategy proposed by \cite{bacchini_2024}. We apply a radiation--reaction (RR) force representing synchrotron cooling \citep[e.g.,][]{Cerutti_2013,Cerutti_2016,Comisso_2021} to the right-hand side of Eq.~\eqref{eq:momenta} for both species during the preturbulent stages, via
\begin{equation}
        \left(\frac{\rmd\bb{u}}{\rmd t}\right)_{\rm RR}= -\kappa \gamma \left[\left(\bb{E} + \frac{\bb{u}}{\gamma}\btimes \bb{B}\right)^2 - \left(\frac{\bb{u}}{\gamma}\bcdot\bb{E}\right)^2 \right] \bb{u}.
    \label{eq:RR_force}
\end{equation}
The radiative cooling is then completely turned off, and in the turbulent stage none of the particle species experience the RR force. The parameter $\kappa$ (different for each species) in Eq.~\eqref{eq:RR_force} is chosen based on the results of nonradiative simulations in such a way that, when the RR force is acting, the total particle energy of each species is kept roughly equal to the initial energy, and that the extra heating going into ions due to reconnection is removed from the system. While in realistic physical systems one would not expect synchrotron cooling to act on (subrelativistic) ions, in this work we simply use this term as a robust way of removing excessive nonthermal energy from the simulation, without necessarily assigning any physical meaning to it. The action of the RR force affects the energy evolution of the particle species, as seen in Fig.~\ref{fig:fields}(d). Excessive ion-energy increase resulting from magnetic-field dissipation is removed until $t=5.6P_0$ (indicated by the vertical dashed line in Fig.~\ref{fig:fields}(d)), at which point we switch off RR forces. From that moment on, ions and electrons start to gain energy only due to the action of MRI turbulence.

The evolution of particle energy distributions, $f(\gamma,t) \equiv \rmd N(\gamma)/\rmd \gamma$, during the turbulent stage (from the moment RR is switched off and until the end of the run) is shown in Fig.~\ref{fig:figure2}(a--b). Ions, as can be seen in panel~(a), significantly heat up, and tend to form a nonthermal power-law tail spanning less than a decade in energy. Electrons (panel~(b)) are efficiently accelerated by MRI turbulence and develop a prominent power-law spectrum.
The electron nonthermal slopes are consistent with previous pair-plasma works \citep{Bacchini_2022,bacchini_2024}, at later stages of the simulation, showing an $f(\gamma)\propto\gamma^{-3}$ scaling (depicted with the dotted line in Fig.~\ref{fig:figure2}(b)) at the end of the run. Both ions and electrons can be accelerated to the point where their Larmor radius becomes comparable to the box size. This leads to the appearance of a high-energy pile-up at the tail of the distribution functions, as observed in other turbulence simulations \citep[e.g.,][]{Zhdankin_2018}, due to the continuous accumulation of energy in our finite-size simulation box.

Turbulent spectra for the poloidal component of the magnetic field $B_{\rm pol} \equiv \sqrt{B_x^2+B_z^2}$ and the toroidal~$B_y$, averaged over two rotational periods $t=10.5P_0\mbox{--}12.5P_0$, are shown in Fig.~\ref{fig:figure2}(c).
Here, perpendicular wavenumbers are taken in the poloidal direction $\bb{k}_\perp = \{k_x,k_z\}$. At fluid scales $k_\perp\langle\rho_i\rangle<1$ (where $\langle\rho_i\rangle$ is the average ion Larmor radius), the spectrum of $B_{\rm pol}$ is roughly proportional to~$k_\perp^{-5/3}$, while the $B_y$ spectrum follows a steeper scaling (for reference, we also show a power law $k_\perp^{-3}$ in the same figure). At ion kinetic scales $k_\perp \langle\rho_i\rangle \approx 1$, both spectral slopes steepen to $\propto k_\perp^{-4}$. This scaling was first reported in PIC simulations by \citealt{zhdankin_etal_2017,Zhdankin_2018}; the spectra steepening is also consistent with previous PIC MRI studies \citep{Bacchini_2022,bacchini_2024}, as well as hybrid simulations \citep{Kunz_2016}, and at large scales it aligns with MHD expectations \citep[e.g.,][]{Walker_2016}. At progressively smaller scales, the slope flattens due to numerical noise.

Angular-momentum transport in MRI turbulence is usually parametrized via an effective viscosity \citep{shakura1973} as 
\begin{equation}
    \langle\alpha\rangle = \frac{1}{\langle p \rangle}\left( \langle M_{xy} \rangle + \langle A_{xy}\rangle + \langle R_{xy} \rangle\right), \label{eq:alpha}
\end{equation}
which is defined through the Maxwell stress $M_{xy} \equiv -B_xB_y/(4\pi)$ and the sum of each $j$-th particle species' anisotropic stress $A_{xy} \equiv -\sum_j (p_{\parallel,j} -p_{\perp,j})B_xB_y/B^2$, where parallel and perpendicular pressure are defined as $p_{\parallel,j} = \mathbf{P}_j\bdbldot\bb{B}\bb{B}/B^2$ and $p_{\perp,j} =\mathbf{P}_j\bdbldot(\mathbb{I} - \bb{B}\bb{B}/B^2)/2 $ in terms of the full pressure tensor $\mathbf{P}_j = m_j\int (\widetilde{\bb{u}}_j\widetilde{\bb{u}}_j/\widetilde{\gamma}_j)f_j(\bb{x},\widetilde{\bb{u}},t){\rm d}^3 \widetilde{\bb{u}}_j$. Here, $f_j(\bb{x},\widetilde{\bb{u}},t)$ is the $j$-th species distribution function, with velocities $\widetilde{\bb{u}}$ computed in a Lorentz-boosted frame moving with the local bulk-flow velocity $\bb{U}_j \equiv (1/n_j)\int \bb{u}_j f_j(\bb{x},\bb{u},t) \rmd^3\bb{u}$. The Reynolds stress is $R_{xy} \equiv \sum_j m_j n_j U_{x,j}U_{y,j}/\Gamma_j$, defined with the bulk Lorentz factor $\Gamma_j \equiv \sqrt{1 + U_j^2 / c^2}$. The angle brackets $\langle ... \rangle$ in Eq.~\eqref{eq:alpha} denote spatial averaging over the simulation box. The evolution of the measured $\langle\alpha\rangle$-parameter, along with its three constituent terms, is given in Fig.~\ref{fig:figure2}(d). We see that, during the turbulent stage, a substantial amount of turbulent viscosity originates from the Maxwell and anisotropic stresses, while the contribution of Reynolds stress is lower by one order of magnitude compared to other stresses closer to the end of the simulation. 
The ordering of the stresses, $\langle R_{xy}\rangle<\langle A_{xy}\rangle\lesssim \langle M_{xy}\rangle$, as seen in Fig~\ref{fig:figure2}(d), is in agreement with previous fluid, hybrid, and (large-scale) pair-plasma PIC simulations \citep[e.g.,][]{Sharma_2006,Kunz_2016,Bacchini_2022,bacchini_2024}, with the ion contribution to stresses being higher than the corresponding electron contribution (see Fig.~\ref{fig:figure2}(e,f)). In particular, the anisotropic stress significantly contributes to the total stress; as shown in previous pair-plasma MRI studies \citep{Bacchini_2022,bacchini_2024}, this contribution could be overestimated due to our necessarily limited system size and could change when modeling regimes with larger scale separation. However, the stress hierarchy we obtain, with $A_{xy}\sim M_{xy}$, is consistent with fluid expectations. The growth of pressure anisotropy is suppressed as most of the ion population (and to a lesser degree electrons) crosses the mirror-instability threshold (see Fig.~\ref{fig:figure2}(g,h)) \citep{kunz_2014,Zhdankin_2021}; therefore, the anisotropic stress is limited by the mirror instability at later stages of the simulation. The Reynolds stress is intermittent in time, and is related to the cycles of channel-flow formation; the ions' contribution to this stress is significantly larger than the electrons'.

\subsection{Electron--ion heating ratio}
\begin{figure}
    \centering \includegraphics[width=0.5\textwidth]{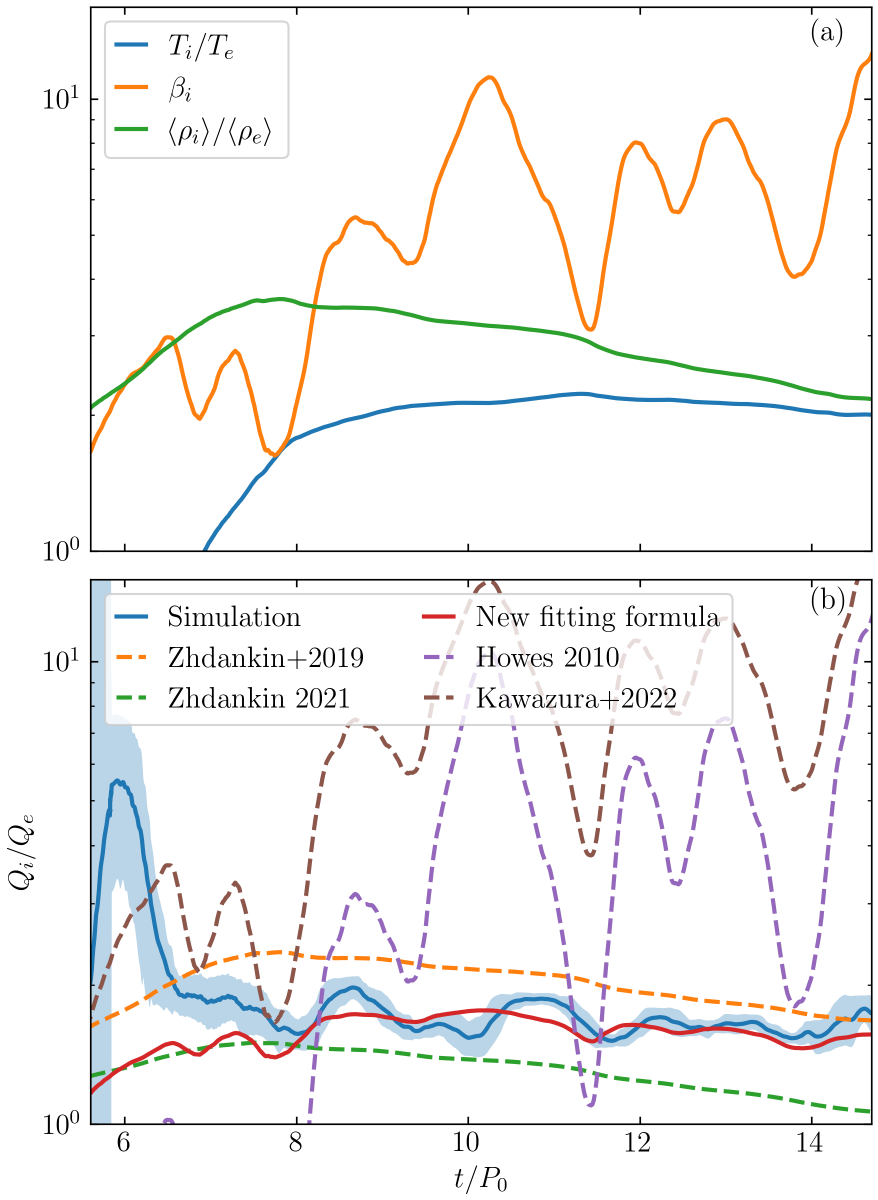}
    \caption{(a) The evolution of dimensionless parameters typically used for heating ratios in the simulation run; (b) Measured heating ratio evolution over time (blue line) versus the heating ratios obtained from \cite{Howes_2010}, \cite{Zhdankin_2019}, \cite{Zhdankin_2021}, and \cite{Kawazura_2022}, along with the heating ratio fitted with our formula, Eq.~\eqref{eq:heatingfit}.}
    \label{fig:heating_ratio_radiative}
\end{figure}

From Eq.~\eqref{eq:momenta}, it immediately follows that the heating rate (or, more accurately, the total energization rate) $Q_j$ for the particle species $j$ can be obtained from the individual particle's Lorentz-factor evolution equation, 
\begin{equation}
   \frac{\rmd \gamma_{j}}{\rmd t} = \frac{q_j}{\gamma_{j} m_j c^2} \bb{E}\bcdot \bb{u}_{j} + s\Omega_0 \frac{u_{j,x} u_{j,y}}{\gamma_{j}c^2}, \label{eq:heating_ratio}
\end{equation}
by summing over all particles $p$, $Q_j \equiv \sum_p m_jc^2 \rmd \gamma_{p,j}/\rmd t$. The evolution of the measured heating ratio $Q_i/Q_e$ over time is shown in Fig.~\ref{fig:heating_ratio_radiative}(b). Following previous works on the topic \citep[e.g.,][]{Howes_2010,Werner_2017,Zhdankin_2019,Zhdankin_2021,Kawazura_2022,satapathy2024}, and to provide a computationally convenient prescription for the heating ratio, we assume that $Q_i/Q_e$ can be expressed as a function of several key dimensionless plasma parameters, such as ion plasma beta $\beta_i=8\pi n_i T_i/B^2$, ion-electron temperature ratio, average Larmor-radius ratio, turbulent viscosity $\langle\alpha\rangle$-parameter, and potentially other quantities: $Q_i/Q_e = f(\beta_i, T_i/T_e, \langle\rho_i
\rangle/\langle\rho_e\rangle,\langle\alpha\rangle,...)$. In this work, we define the temperature of ions and electrons via average the Lorentz factor $\langle\gamma_j\rangle = \kappa_{23}(1/\theta_j) - \theta_j$, where $\kappa_{32}(x) = \mathcal{K}_3(x)/\mathcal{K}_2(x)$, and $\mathcal{K}_3(x)$, $\mathcal{K}_2(x)$ are modified Bessel functions of the second kind of order 3 and~2, respectively.

The temporal evolution of some of the aforementioned parameters in the turbulent stage is shown in Fig.~\ref{fig:heating_ratio_radiative}(a). Comparing the two panels of Fig.~\ref{fig:heating_ratio_radiative}(a,b), we see that the heating ratio shows a relatively weak dependence on~$\beta_i$, in contrast to the results by \cite{Howes_2010} and \cite{Kawazura_2022}, which would predict significant oscillations in the heating ratio correlated with $\beta_i$ (depicted in Fig.~\ref{fig:heating_ratio_radiative}(b) for comparison). {These large-amplitude oscillations in the heating ratio implied by the prescriptions were not noted in previous works \citep[e.g.][]{Kawazura_2019,kawazura_etal_2020,Kawazura_2022}, which may be due to the implicit time averaging of parameters in those heating-ratio formulas. We also note that the heating ratio measured in this work is generally lower than other prescriptions, potentially because in our semirelativistic regime, the effective (relativistic) mass ratio is lower than the rest-mass ratio employed by \cite{Howes_2010} and \cite{Kawazura_2022}.  Additionally, a significant pressure anisotropy in our simulation can act as a different energization driver in accretion flows \citep{Sharma_2007}, whereas the focus of earlier works\footnote{ {Note that in \citep{Kunz_2018} the pressure-anisotropy influence on plasma heating was carefully examined. However, a direct comparison of their results with the present work would be challenging, as here a significant fraction of plasma exceeds the mirror-instability threshold, violating the assumptions of their analysis.}} \citep{Howes_2010,Kawazura_2022} was on different routes of dissipation through Alfv\'enic and/or compressive fluctuations, neglecting pressure anisotropy.}  In our simulations, the heating ratio correlates most strongly with the Larmor radius ratio~$\langle\rho_i\rangle/\langle\rho_e\rangle$, with weaker oscillations correlating to~$\beta_i$. The heating ratios from \cite{Zhdankin_2019} and \cite{Zhdankin_2021} (also shown in Fig.~\ref{fig:heating_ratio_radiative}(b)) give a good overall agreement with our measured heating ratio. This applies particularly to the \cite{Zhdankin_2021} scaling $Q_i/Q_e\propto (\langle\rho_i\rangle/\langle\rho_e\rangle)^{1/3}$ for solenoidal driving. This is not surprising, as MRI turbulence is mostly solenoidal \citep[e.g.,][]{bacchini_2024}. We expect that at larger scale separation, as the driving scale ($k_{\rm MRI}$) is pushed further away from the kinetic scales, the power of compressive fluctuations of the ions will further decrease, as compressive fluctuations tend to localize at larger scales \citep{Zhdankin_2021}, and kinetic-scale turbulence would inevitably excite compressive modes. By including a weak dependence on $\beta_i$ in this formula, with the functional form introduced in previous studies \citep[e.g.,][]{Quataert_1998,Howes_2010,kawazura_etal_2020},
we can obtain better agreement with the measured heating ratio. Therefore, in the case of MRI-driven turbulence in typical collisionless accretion disks at large $\beta_i\approx10$, we propose that the heating ratio can be estimated as    
\begin{align}
    \frac{Q_i}{Q_e} =  1.3\left(\frac{\langle\rho_i\rangle}{\langle\rho_e\rangle}\right)^{1/3}\exp\left(-\frac{1}{2\beta_i}\right).
    \label{eq:heatingfit}
\end{align}
The proposed formula is shown in Fig.~\ref{fig:heating_ratio_radiative}(b). It can be seen that Eq.~\eqref{eq:heatingfit} provides a very good approximation for the heating ratio measured in the simulation during statistically steady-state MRI turbulence (starting from $t\approx8 P_0$), where the system parameters do not change rapidly (see also the discussion in Appendix~\ref{sec:appA}). In order to generalize this formula to the regime when both ions and electrons are relativistic, where one would expect $Q_i/Q_e\rightarrow 1 $, we would need to conduct a larger parameter study, which is beyond the scope of this work and will be pursued in the future. 

\section{Discussion}
In this work, we have presented the first simulations of fully kinetic, three-dimensional, MRI-driven turbulence in semirelativistic ion--electron plasma with a realistic proton-to-electron mass ratio. These runs dramatically surpass in size and quality all previous similar works (including our own). Such large simulations are unavoidable if one wishes to study the properties of turbulence in the realistic proton--electron plasma found in RIAFs. 
In this sense, our work allows us to model MRI turbulence in the most realistic physical conditions attained so far.  {The largest scale separation $\omega_{\mathrm{c},i}/\Omega_0=8$ employed here is dictated by computational limitations and by the complexity of fully kinetic simulations. Real SMBH accretion flows are characterized a much larger (by orders of magnitude) scale separation and can be qualitatively different in some aspects (e.g., pressure-anisotropy enhancement and channel-flow formation in the turbulent regime). However, since our results agree well with both previous fluid and kinetic studies, we are confident that our approach provides a promising foundation for further exploration of RIAFs with first-principles calculations.}

Our largest simulation shows a temporal evolution qualitatively comparable to previous pair-plasma simulations and consistent with fluid and hybrid models. At the linear stage, we observe the formation of channel flows, which are then disrupted by the large-scale reconnection, after which the system enters a nonlinear, turbulent regime.  {In this regime, the channel flows can reform and get disrupted again, which can contribute to plasma heating. Previous pair-plasma studies \citep[e.g.,][]{Bacchini_2022,bacchini_2024} suggest that this cycle is a feature of smaller simulations, and tends to disappear for larger scale separations. Thus, the recurring large-scale reconnection may have a more pronounced impact on heating in our modest scale-separation runs.}

To suppress the transfer of magnetic energy (e.g.,\ via reconnection) to particles at the end of the linear MRI growth stage, before MRI turbulence is fully developed, we have a employed radiation-reaction force acting on both ions and electrons. We turn off the RR force when the turbulence is fully developed. This approach significantly reduces the impact of initial conditions on the system. 
Such a strategy was successful in previous work \citep{bacchini_2024}, and here it allows us to start the turbulent stage from a ``clean'', quasithermal ion distribution that is not polluted by large-scale (system-size) reconnection events prior to the development of nonlinear MRI turbulence. In this way, we have shown for the first time (and agnostically of initial conditions) that MRI turbulence can accelerate not only electrons but also ions to nonthermal distributions. This agrees with the solenoidal-driving investigation by \citep{Zhdankin_2021}, as one expects MRI-driven turbulence to be mostly solenoidal \citep{gong2020, bacchini_2024}.
Turbulent magnetic spectra, even at the modest scale separation (between the MRI injection scale and kinetic scales) employed in these runs, feature MHD-like spectral slopes at fluid scales, as well as a spectral break and a steepening of the spectra when reaching kinetic scales, in agreement with all previous works on MRI.
Our analysis of angular-momentum transport is also in agreement with hybrid-kinetic \citep{Kunz_2016} and Braginskii-MHD \citep[e.g.,][]{Sharma_2006} MRI simulations, as well as previous PIC MRI studies \citep{Bacchini_2022,bacchini_2024}.
Even with our small fluid-to-kinetic scale separation, the stress ordering we have obtained is reassuring that the simulation provides (at least qualitatively) correct results. We have found that significant AMT is promoted by pressure anisotropy, which is regulated by the action of mirror instabilities. For the first time, we have been able to quantify the individual impact of different particle species on the~AMT. The ion anisotropic stress is slightly larger (on average) than that of the electrons, which is related to the fact that the (approximate) mirror instability threshold for electrons is lower than that for ions \citep{Zhdankin_2021}. 

Our main result is the analysis of the energy partition between ions and electrons in a fully kinetic MRI simulation. This is particularly important for the global modeling of astrophysical accretion flows around SMBH, such as those targeted by the~EHT, which necessarily employs subgrid kinetic prescriptions in fluid simulations. The simple heating prescription we provide in Eq.~\eqref{eq:heatingfit} can be used to describe electron and ion thermodynamics in GRMHD models of SMBH accretion flows \citep[e.g.,][]{Ressler_2015,Dexter_2020,Scepi_2022,Scepi_2023}. 
Compared to previous works (e.g.,\ \citealt{Howes_2010,Kawazura_2022}), our measured heating ratio shows a weaker dependence on the plasma beta. Similar behavior was observed in hybrid-kinetic simulations \citep{arzamasskiy_etal_2023}. We found that a simple formula (our Eq.~\eqref{eq:heatingfit}) based on the local ion-to-electron Larmor-radius ratio provides a good fit to the measured heating ratio for the parameters of interest for RIAFs. The power index $p$ in this fitting formula, determined by the type of driving, is set to be $p = 1/3$, in line with previous works \citep{Zhdankin_2021}.
We note that this heating ratio does not agree with the energy partitions obtained from gyrokinetic studies \citep{Howes_2010,Kawazura_2019}, and inferred in reduced-MHD simulations \citep{Kawazura_2022,satapathy2024}. 
This may be related to qualitative differences between those models and our approach. For example, the fact that we study semirelativistic plasma, which is realistic for RIAFs and entails a significant fraction of particles experiencing NTPA, diverges substantially from the gyrokinetic and reduced-MHD assumptions. In addition, the net magnetic flux in our simulation is weak and directed along the $z$-direction, in contrast to the cited works where a strong mean toroidal field is present instead. 

We believe that this work represents a significant step forward in improving our understanding of the microscale physics of plasmas around SMBHs, and sets a new standard for future fully kinetic simulation studies as well as for subgrid models employed in global GRMHD simulations of plasma accretion. 
This study demonstrates the feasibility of first-principles kinetic modeling of collisionless accretion flows, paving the way for more precise and realistic simulations in the future.

\begin{acknowledgments}
E.A.G.\ would like to thank M.\ Hoshino, L.\ Sironi, A.\ Philippov, J.\ Stone and J.\ Davelaar for useful discussions throughout the development of this work.
F.B.\ acknowledges support from the FED-tWIN programme (profile Prf-2020-004, project ``ENERGY'') issued by BELSPO, and from the FWO Junior Research Project G020224N granted by the Research Foundation -- Flanders (FWO). 
V.Z.\ acknowledges support from NSF grant PHY-2409316.
We also acknowledge support from NASA ATP grants 80NSSC22K0826 and 80NSSC22K0828, and National Science Foundation grant AST 1903335 to the University of Colorado.
The computational resources and services used in this work were partially provided by the VSC (Flemish Supercomputer Center), funded by the Research Foundation Flanders (FWO) and the Flemish Government – department EWI. We acknowledge LUMI-BE for awarding this project access to the LUMI supercomputer, owned by the EuroHPC Joint Undertaking, hosted by CSC (Finland) and the LUMI consortium through a LUMI-BE Regular Access call. The authors also acknowledge the Texas Advanced Computing Center (TACC) at The University of Texas at Austin for providing computational resources that have contributed to the research results reported within this paper. URL: http://www.tacc.utexas.edu. 
\end{acknowledgments}

\appendix
\section{Heating ratio time dependence}\label{sec:appA}
\begin{figure}[h]
    \centering \includegraphics[width=0.5\textwidth]{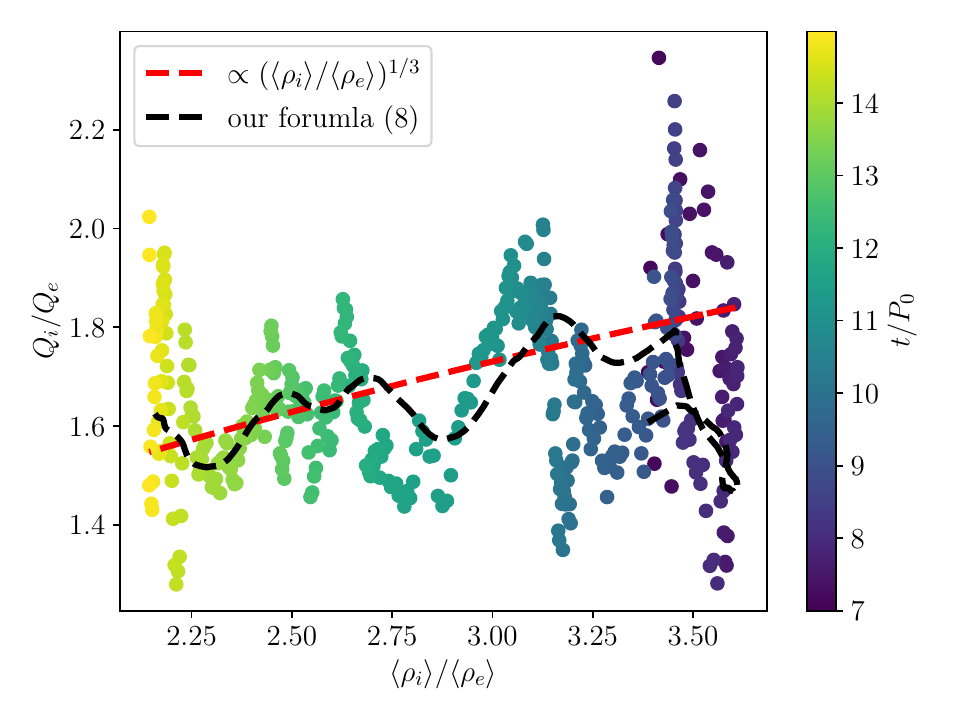}
    \caption{Scatter plot of $Q_i/Q_e$ vs.\ $\langle\rho_i\rangle/\langle\rho_e\rangle$. The later times of the simulation are shown in lighter colors. The heating ratio proposed in this work is shown by the black dashed line, and the red solid line represents \cite{Zhdankin_2021}'s formula for the heating ratio with solenoidal driving.}\label{fig:heating_ratio_scatter}
\end{figure}

In this work, we treat each measurement at each time step as a separate data point. We acknowledge the fact that one would expect that it might introduce an explicit time dependence to the heating ratio, i.e., $Q_i/Q_e = f(t,T_i/T_e,\beta_i, \langle\rho_i
\rangle/\langle\rho_e\rangle,\langle\alpha\rangle,...)$, which can impact the result. While this can be true, we note that the previous PIC works \citep{Zhdankin_2019,Zhdankin_2021}, and all PIC studies in general, in the absence of a mechanism that would remove continuous growth of energy in a closed simulation box, potentially suffer from such a problem. In order to address this problem, we provide a scatter plot of $Q_i/Q_e$ as a function of $\langle\rho_i
\rangle/\langle\rho_e\rangle$ (see Fig.~\ref{fig:heating_ratio_scatter}). There, we also show the proposed heating ratio \eqref{eq:heatingfit}, as well as the formula given in \citealt{Zhdankin_2021} for solenoidal driving. We note that our approximation \eqref{eq:heatingfit} provides very good agreement with the measurements, which implies that the approach of treating different time steps as separate data points is valid. The heating-ratio variations are due to the change of other parameters (in our case, $\beta_i$), and may also be impacted by the noisy nature of PIC simulations. More accurate data measurements would require more (and probably larger) simulations, which is beyond the scope of this work.

\FloatBarrier
\bibliographystyle{aasjournal.bst}
\bibliography{literature}{}

\end{document}